# Approaching the theoretical limit of wicking on textured surfaces

**Authors:** Zhaoyang Lv[1]†, Chen Ma[2]†, Li-Chen Huang[1], Yanshen Li[1,3*]

**Affiliations:**

[1]School of Engineering Science, University of Chinese Academy of Sciences, Beijing 101408, PR China

[2]The Hong Kong Polytechnic University, Hong Kong SAR 999077, PR China

[3]State Key Laboratory of Nonlinear Mechanics, Chinese Academy of Sciences, Beijing, 100190, PR China

*Corresponding author. Email: liyanshen@ucas.ac.cn

†These authors contributed equally to this work

**Abstract:** Wicking in a capillary tube could happen as long as the liquid contact angle is smaller than 90°, making it possible for weak hydrophilic liquids to spontaneously invade the tube. For textured surfaces, energy minimization argument predicts the same. However, wicking of weak hydrophilic liquids on textured surfaces has not been possible due to energy barriers induced by the textures. We demonstrate how these barriers could be avoided by adjusting the shape and arrangement of the pillars, thus the wettability required for wicking approaches the theoretical limit. An unprecedented wicking contact angle of 82° is reported. More surprisingly, wicking coefficients of such surfaces can be larger than that of rectangular grooves at the same porosity. These findings may significantly advance biomedical and thermal management technologies.

Wicking is the spontaneous invasion of a porous material by liquids. It is ubiquitous in everyday life, and the best-known example is the imbibition of liquids into capillary tubes[1–4]. Wicking is also widely utilized in industrial applications, including thermal management[5–8], biomedical devices[9–11], food[12] and textile[13,14] industries. In many of these applications, the porous material takes the form of quasi-two-dimensional (quasi-2D) structures like textured surfaces. In capillary tubes, wicking can occur as long as the tube wall is hydrophilic, i.e., when the liquid contact angle $\theta < 90°$[3,4]. On textured surfaces, however, wicking occurs only at much smaller contact angles: to date, the largest reported values are 56° for surface textured with vertical pillars[15] and 63° for deflected ones[16]. This significantly limits the variety of usable liquid–solid pairs, forcing wicking applications to rely on high-surface-energy coatings or low-surface-tension liquids, each of which brings its own limitations.

The minimization of surface energy, or the so-called "thermodynamic" argument, predicts that wicking can happen when the liquid contact angle $\theta$ is smaller than a critical value $\theta_{\mathrm{cr}}^{E}$. For textured surfaces, the critical contact angle $\theta_{\mathrm{cr}}^{E}$ approaches 90° if the roughness of the texture is large enough[17], for example, when the surface is textured with very tall and thin pillars. But the disconnected nature of the surface roughness induces energy barriers that hinder wicking: the liquid does not "know" the existence of the next row of pillars[17]. To account for this, it was proposed to assume the liquid front between two rows of pillars takes the shape of a one-dimensional (1D) prism[17,18], so that the liquid continues wicking when it touches the next row of pillars. This 1D prism model also predicts a critical contact angle $\theta_{\mathrm{cr}}^{W}$ that approaches 90°. However, in reality, the real $\theta_{\mathrm{cr}}$ can hardly go beyond 60°[15,19] due to all sources of energy barriers[19–21]. The shape of the meniscus at the liquid front is complex because the liquid needs to maintain its contact angle $\theta$ with all the solid walls in the texture. Therefore, it is not easy to completely figure out all the sources of energy barriers. To make things worse, contact angle hysteresis and pinning on sharp edges of non-circular pillars adds more (potential) sources to energy barriers. Thus, it might seem impossible to avoid all possible energy barriers and make $\theta_{\mathrm{cr}}$ approach the thermodynamic limit $\theta_{\mathrm{cr}}^{E}$ (note $\theta_{\mathrm{cr}}^{E}$ cannot be reached because it assumes the liquid front being rectangular, which is not real). Here in this paper, we demonstrate that the energy barriers are dictated by two key processes and can be circumvented by simply adjusting the shape and arrangement of the pillars. The thermodynamic limit $\theta_{\mathrm{cr}}^{E}$ is approached with the difference $\theta_{\mathrm{cr}}^{E} - \theta_{\mathrm{cr}}$ being smaller than 2° and an unprecedented wicking contact angle of 82° is reported. We also demonstrate how these findings could provide new solutions for biomedical applications and boost the heat transfer rate of liquid-vapor phase change technologies.

## Local energy barriers and the way to avoid them

Figure 1A shows a direct comparison of the critical contact angles of capillary tubes, grooves and textured surfaces. For continuous structures such as capillary tubes or grooves, there is no local energy barrier that hinder wicking so that $\theta_{\mathrm{cr}} = \theta_{\mathrm{cr}}^{E}$. For square grooves, $\cos\theta_{\mathrm{cr}} = w/(w+2h)$, where $w$ and $h$ are the width and height of the groove. Obviously, $\theta_{\mathrm{cr}}$ approaches 90° when $w/h$ is large enough. For textured surfaces, surface energy minimization gives $\cos\theta_{\mathrm{cr}}^{E} = (1-f)/(r-f)$ where $f$ is the projected area fraction of the pillars and $r$ the solid roughness (ratio of the actual solid area over its projected one). It is easily found that $\theta_{\mathrm{cr}}^{E}$ approaches 90° when the roughness $r \to \infty$. However, due to local energy barriers, the achievable critical contact angle $\theta_{\mathrm{cr}}$ for surfaces textured with square pillars does not exceed 50° (See SM).

To better study the shape of the liquid front and the corresponding surface energy of the system, wicking experiments were performed on different textured surfaces and the energy landscapes during wicking were obtained by Surface Evolver (SE) simulation. To better observe the shape of the menisci, four different kinds of surfaces with submillimeter textures were made on copper by milling. The pillar heights $h$ is all kept 0.95 mm and typical pillar width $a$ all kept around 0.2 mm. To avoid the influence of gravity, the samples were put horizontally to see if wicking can happen (see SM for details).

On these textured surfaces, we have identified two key processes during wicking that control the energy barriers: the first is whether the liquid could engulf the outmost row of pillars, the second is whether the liquid could touch the next row of pillars. Take the commonly used surface textured with regularly arranged square pillars as an example, when the advancing angle is $\theta_a = 55 \pm 3°$ (Fig. 1D where $a = b$ and the pillar height $h/a = 4.75$), the liquid stops at the rightmost surfaces of the pillars in the outmost row. To engulf the base of this outmost row of pillars, the two menisci on the two sides of one pillar need to advance and merge with each other, we call this the "pre-merge" process. However, the total surface energy of the system in this pre-merge process increases, see the red line in Fig. B for the energy landscape obtained by Surface Evolver (SE) simulation with the same parameters as in the experiments (because the textured surface is periodic in the wicking direction, the evolution of the shape of the liquid front is also periodic, thus the energy landscape in one period should suffice). Lowering the contact angle to about 49° makes this energy barrier disappear (black line). Let $\theta_{\text{merge}}$ be the critical contact angle below which the two menisci could merge spontaneously without any energy barrier, more wicking experiments and SE simulations spanning the parameter space ($f$, $\theta$, $h/a$) on square-pillar textured surfaces show that $\theta_{\text{merge}}$ first increases with $f$ and then decreases slowly, yielding a maximum value of $\approx 50°$. Interestingly, this value is almost independent of the aspect ratio $h/a$. That is to say, the maximum value of $\theta_{\text{merge}}$ is mainly dictated by the shape of the pillar. Thus, to further decrease the energy barrier induced by the engulf process, or in other words, to increase $\theta_{\text{merge}}$, one must change the shape of the pillars.

Speaking of the shape, the outmost face of the square pillars in the outmost row is perpendicular to the wicking direction. Thus, the two menisci must propagate in a direction perpendicular to the wicking direction to merge with each other, this is rather difficult. It will be easier for the two menisci to merge if they can propagate in a direction more "parallel" to the wicking direction, this could be done by making the pillars pointy in the wicking direction, for example, a diamond cross-sectional shape. Fig. 1E shows a surface textured with regularly arranged diamond pillars with a half tip angle $\alpha = 10°$ and the pinned position of the liquid front when the advancing angle is $\theta_a = 74 \pm 3°$. Notice that the liquid has fully engulfed the base of the outmost row of pillars, meaning $\theta_{\text{merge}} > 74°$. Actually, SE simulation shows that $\theta_{\text{merge}}$ of such slim-diamond shaped pillars is 75°, much larger than that of square pillars. The liquid front on this surface stops before it touches the next row of pillars, meaning this "pre-touch" process also contributes to the local energy barrier during wicking. Notice that this process is what the 1D prism model tries to describe. The red line in Fig. 1C shows the energy landscape on this surface when the contact angle $\theta$ is higher at 81°. The two energy barriers cause by the "pre-merge" process and the "pre-touch" process are prominent and there is no other energy barrier. This confirms our main finding that the local energy barriers are dictated by two key processes: the "pre-merge" process and the "pre-touch" process. Let $\theta_{\text{touch}}$ be the critical contact angle below which the liquid could touch the next row of pillars. Then the real critical contact angle for wicking should be $\theta_{\text{cr}} = \min\{\theta_{\text{merge}}, \theta_{\text{touch}}\}$.

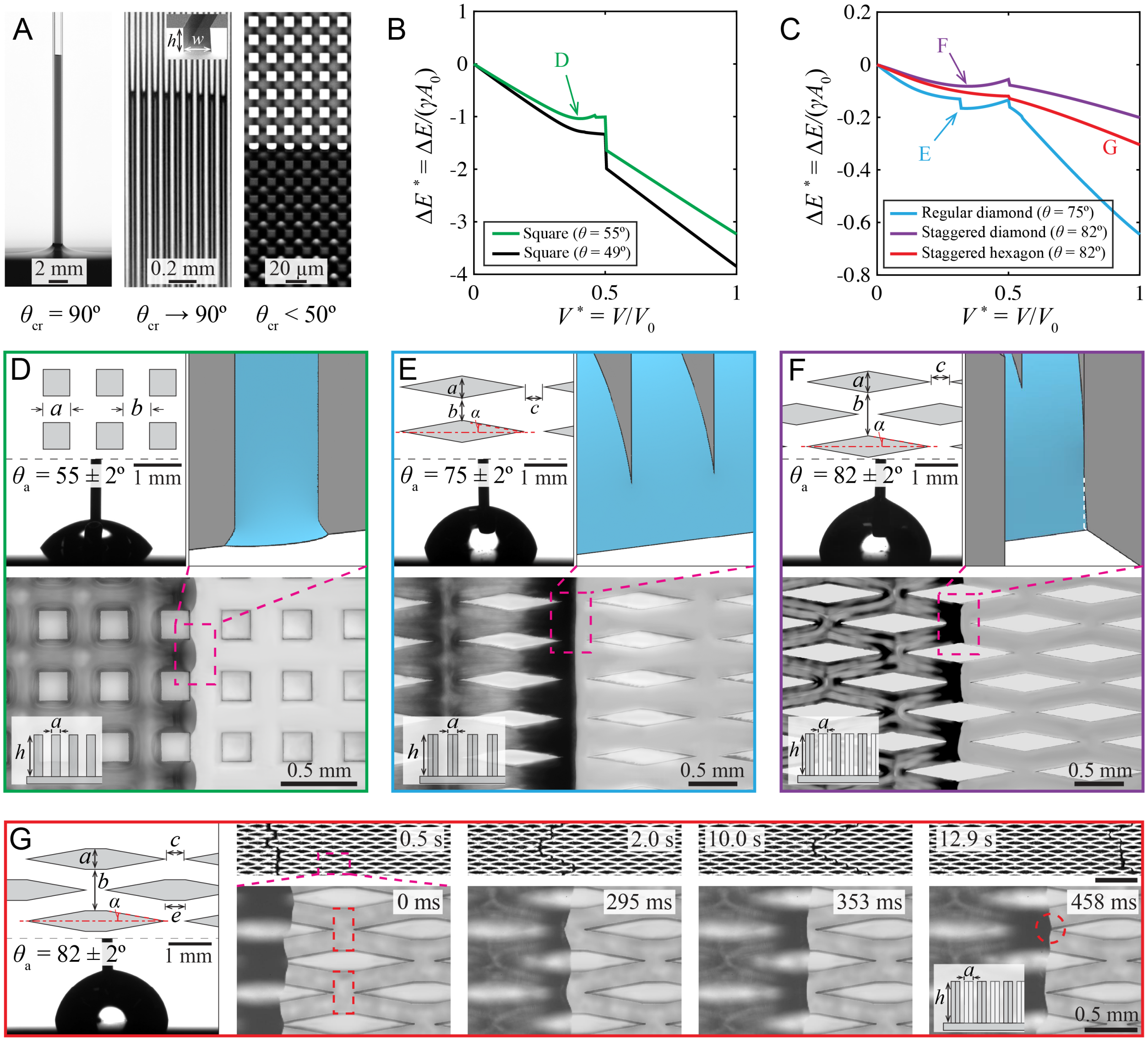

Fig. 1. Energy barriers on surfaces textured with straight pillars and the way to avoid it. A. Snapshots of wicking in three typical structures and their wicking critical contact angle $\theta_{cr}$. B-C. Energy landscapes of the wicking process for different contact angles on four different pillar-textured surfaces: regularly arranged square, regularly arranged diamond, staggered diamond and staggered hexagon. The wicking process is periodic in the wicking direction so the energy landscape in one period is shown. The geometrical parameters of the four surfaces are the same as that in D-G. D-E. Shape and position of the static liquid front on the regularly arranged square/diamond pillar textured surfaces at different contact advancing contact angles $\theta_a$. F. The wicking process on staggered diamond pillar textured surface for $\theta_a = 77 \pm 3°$ and the evolution details of the liquid front. The half tip angle of the diamond is $\alpha = 10°$. Wicking in the red dashed region is the most difficult portion because it is a diverging groove. G. The wicking process on staggered hexagon pillar textured surface and the evolution details of the liquid front. The pre-touch regions outlined by the red dashed boxes are parallel grooves. This further avoids energy barrier and an unprecedented wicking contact angle $\theta_a = 82 \pm 3°$ is found. Unlabeled scale bars are 2 mm.

To reduce the local energy barrier caused by the pre-merge process, i.e., to increase $\theta_{touch}$, one can simply rearrange the pillars in a staggered manner, see Fig. 1F for a sketch of the pillar arrangements. The snapshots of the wicking process for a liquid advancing contact angle $\theta_a = 77 \pm 3°$ is also shown. SE simulation predicts a $\theta_{cr} = 79°$, consistent with the experiments, yet this value is still smaller than the thermodynamic limit $\theta_{cr}^{E} = 83°$. The reason is, in the pre-touch process, the liquid front must propagate in a diverging square groove, i.e., the groove is becoming wider, see the dashed region in Fig. 1F. This is slightly harder than propagating in a parallel square groove, thus the energy barrier caused by the pre-merge process can still be seen if $\theta = 81°$, see the blue line in Fig. 1C.

To remove this energy barrier, one can further change the "pre-touch region" to a parallel groove by changing the shape of the pillar to a hexagon, see Fig. 1G. Obviously, the length of this parallel region $e$ should be equal to the pillar distance $c$. Consequently, the liquid paths on such a surface is only composed of parallel grooves, thus providing the largest possible wicking contact angle. An unprecedented wicking contact angle as high as 83 ± 3° was observed, also no energy barrier is observed in the energy landscape (black line for $\theta = 81°$ in Fig. 1C). Notice that on this surface, the thermodynamics limit for wicking is $\theta_{cr}^{E} = 83.9°$. The different between the measured value and the thermodynamics limit is smaller than 2° and is undistinguishable in experiments, thus we consider the critical contact angle for wicking on such surfaces have reached the thermodynamic limit.

To further compare the critical contact angle for wicking on these new textured surfaces with the thermodynamic limit, surfaces textured with slim-diamond pillars and slim-hexagon pillars with different surface fraction $f$ but the same $a$ and $h$ are fabricated and tested for wicking. The results are shown in Fig. 2. The thermodynamic limit for surfaces textured with vertical pillars is reorganized into the following form

$$\cos\theta_{cr}^{E} = \frac{1-f}{1+(4\frac{h}{a_e}-1)f} \tag{1}$$

so that $\theta_{cr}^{E}$ is a function of $f$ and $h/a_e$, where $a_e$ is the hydraulic diameter of the cross-sectional shape of the pillars. See SM for the hydraulic diameters of different shapes. Two special cases are circular and square shapes whose hydraulic diameters equals to their diameter and side length, respectively. From Eq. (1), it is easily found that the larger $h/a_e$, the larger $\theta_{cr}^{E}$.

Previously reported wicking contact angles on different textured surfaces are also shown[5,15–19,22–33]. It can be seen that the largest wicking contact angle for surfaces textured with vertical pillars is 56.3° for circular pillars with a quite large aspect ratio $h/a = 6.7$[15]. To further increase the wicking contact angle, deflected[16] or tilted[19] pillars are used and only unidirectional wicking were found for contact angles 60° and 63°. Even so, these values are still quite far from the thermodynamic limit (assuming the pillars are vertical). However, on the newly designed surfaces textured with slim-hexagon pillars, the wicking contact angle is very close to that of the thermodynamics limit with a difference smaller than 2°. Notice that wicking on the newly designed surfaces is not unidirectional or bidirectional but can happen anisotropically in all directions. In addition, while the previous high wicking contact angles were only achieved with very sparse pillars (small $f$), our new textured surfaces can have very large wicking contact angle at all pillar densities. Especially, it was normally believed that dense pillars will induce more pinning sites and consequently stronger energy barriers, but our new results indicate otherwise if the shape and arrangement of the pillars are chosen wisely.

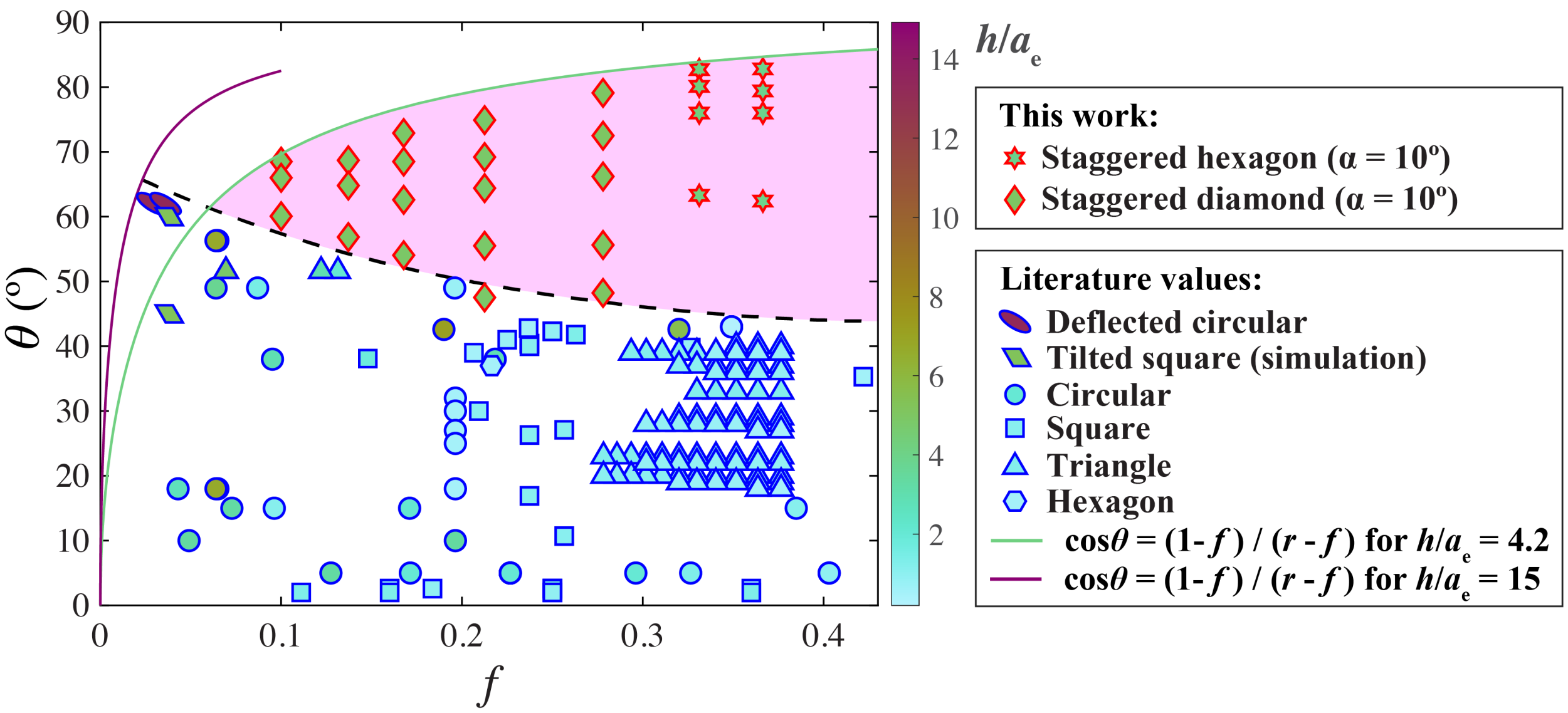

Fig. 2. Wicking contact angles $\theta$ on different pillar textured surfaces characterized by the surface fraction $f$. Different symbols stand for different shape (and arrangement) of the pillars. The solid lines are the thermodynamics limit for different aspect ratio $h/a_e$ of the pillars, where $h$ is the pillar height and $a_e$ the hydraulic diameter of the pillar cross-section. Both the literature values[5,15–19,22–33] and the results of this work are plotted. Note that most of the literature values did not specify if the contact angle is the Young's contact angle. But for the results of this work, the advancing contact angle $\theta_a$ is used. All the previous results are below the black dashed line, while the results on the new textured surfaces in this work can go beyond and reach the thermodynamic limit (the pink region).

**New solutions for biomedical and thermal management applications – making water wicking on textured metal surfaces.**

Increasing the critical wicking contact angle $\theta_{cr}$ to the thermodynamics limit makes a large variety of liquid-solid pairs possible in wicking related applications, even without the aid of surface coatings. For example, water has a contact angle $\theta > 60°$ on many commonly used metal surfaces[34], including titanium, copper, silver and iron, etc. Among them, the titanium alloy Ti-6Al-4V, which contains 6% aluminum and 4% vanadium, was widely used in aerospace due to its high strength-to-weight ratio and exceptional corrosion resistance. It is also widely used as orthopedic and dental implants in biomedical applications due to its good biocompatibility[35]. The wettability of Ti-6Al-4V implants significantly influences the osseointegration duration: hydrophilic Ti-6Al-4V implants can reduce the osseointegration period from 3-6 months to 4-8 weeks in some dental applications, enabling earlier loading. However, as-polished Ti-6Al-4V typically has a water contact angle around 60-80°[35], which is a moderate to weak hydrophilicity. Existing approaches to making Ti–6Al–4V surfaces hydrophilic typically combine micro/nanoscale roughness with hydrophilic surface chemistry. However, this chemistry—usually provided by surface hydroxyl (–OH) groups—is metastable and ages over time, requiring careful storage and handling. With the critical contact angle for wicking now raised to 82° on our surface, Ti–6Al–4V can instead be made strongly hydrophilic through wicking alone, without any chemical surface modification.

Fig. 3 shows a direct comparison of the wetting state of a 9 µL water drop on Ti-6Al-4V surfaces textured with regular square pillars (Fig. 3A-C) and staggered slim hexagon pillars (Fig. 3D-F).

The two textured surfaces have the same geometrical parameters as in that of Fig. 1D and G, except that the pillar height is $h = 0.7$ mm. Their critical contact angles for wicking are ~ 50° and 79°, respectively. The samples are only ultrasonically cleaned sequentially in acetone, ethanol and de-ionized (DI) water before the wettability test. The advancing contact angle on smooth Ti-6Al-4V after the cleaning procedure is $\theta_{cr} \approx 72°$. As expected, on the first sample, the water drop is in Wenzel state with a very large apparent contact angle $\theta^* \approx 120°$. On the second sample, on the contrary, water imbibes within the pillars, forming a film. After water has taken up all the spaces within the pillars, an extra 9 µL water drop was deposited. Not surprisingly, the drop is now in Cassie state with an apparent contact angle $\theta^*$ ranging from 19° (in the long axis) to 30° (short axis). This proves that wicking can happen on the widely used Ti-6Al-4V alloy which is normally moderately/weakly hydrophilic, thus making it strongly hydrophilic only with surface texture., without the aid of surface chemical treatment.

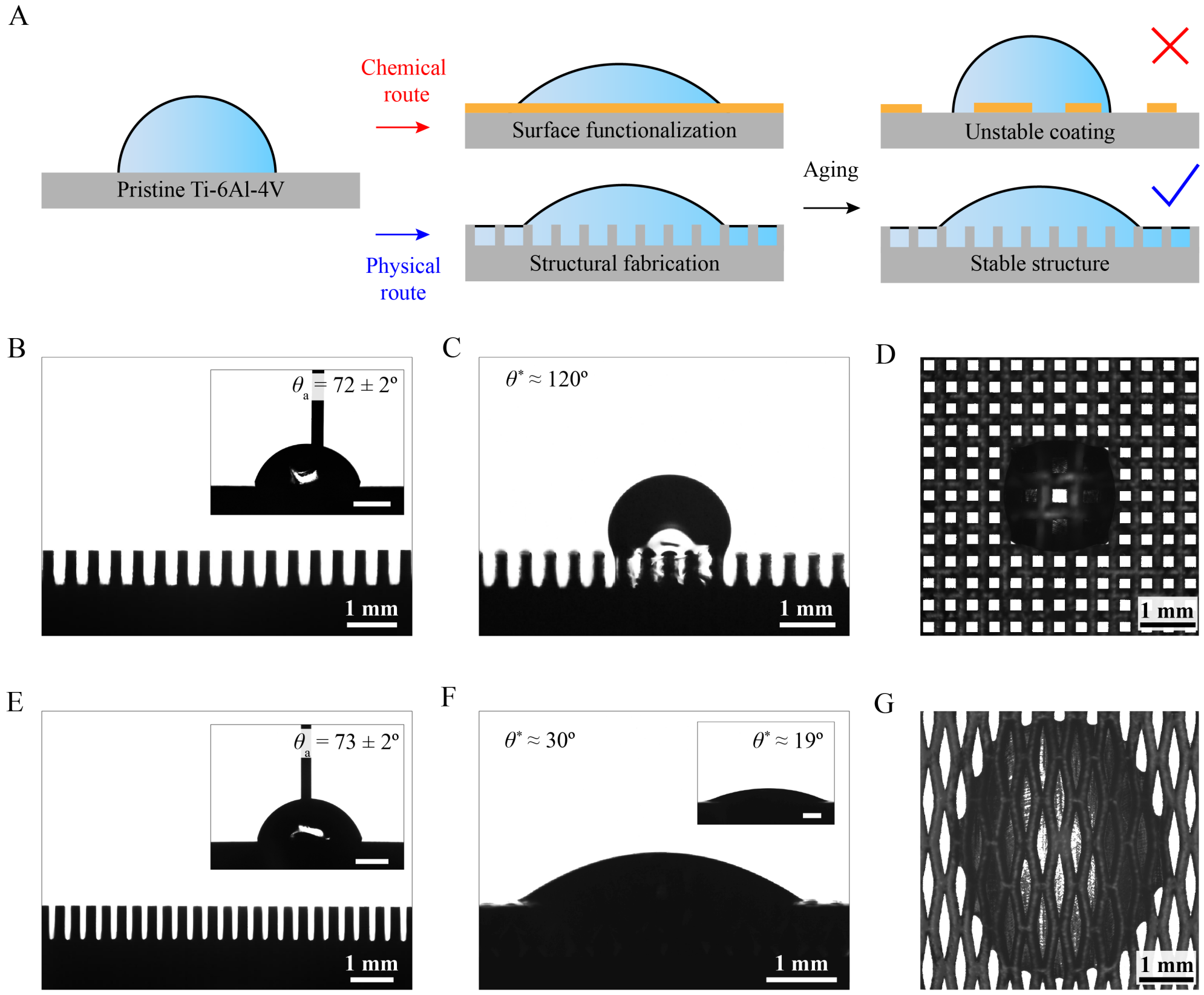

Fig. 3. Wetting states of a 9 µL water drop on two differently textured surfaces both made of the Titanium alloy Ti-6Al-4V. A. Schematic of the durability against aging: chemical coatings vs. physical routes. Side views of the surfaces textured with regular square pillars (B) or staggered hexagon pillars (E). Insets shown the advancing contact angles on a flat surface that has the same material and roughness as the textured surface. C-D. Side view and top view of a 9 µL water deposited on the square pillar textured surface, which is in the Wenzel state. The apparent contact angle is about 120°. F-G. Side views and top view of a 9 µL water drop deposited on the hexagon pillar textured surface. The drop is in the Cassie state. The apparent contact angles are 30° and 19° in the two perpendicular directions. All scale bars are 1 mm.

While Ti-6Al-4V is widely used in biomedical applications, copper is widely used in thermal management technologies, such as in heat pipes and in boiling heat exchangers[5,36]. Water is usually used in pair with copper in these technologies, forming the most common water-copper pair (other pairs include ammonia-aluminum and sodium-steel, etc.). Wicking of water in porous copper material is essential in heat pipes and works best in boiling heat exchangers to enhance the critical heat flux (CHF). However, the static contact angle of air-exposed copper is often greater than 70°[5], making it impossible for wicking. Luckily, a thin layer of CuO is easily formed on the copper surface and its contact angle with water is 20-40°[5], making wicking possible. However, the thermal conductance of CuO is only ≈ 18 W/(m·K) and that of $Cu_2O$ is even smaller, both much smaller than that of Cu which is ≈ 401 W/(m·K). Though normally the oxidized copper layer is very thin, sometimes it is still unwanted and a thin layer of Ag is deposited to prevent oxidation[37]. In this scenario, wicking is again not possible on traditionally textured Ag surfaces. However, with the new surface texture methodologies introduced in this work, wicking on textured Ag surface should be possible, thus one could further increase the heat transfer rate of these technologies.

## Wicking speeds of different surfaces and the wicking performance with phase change

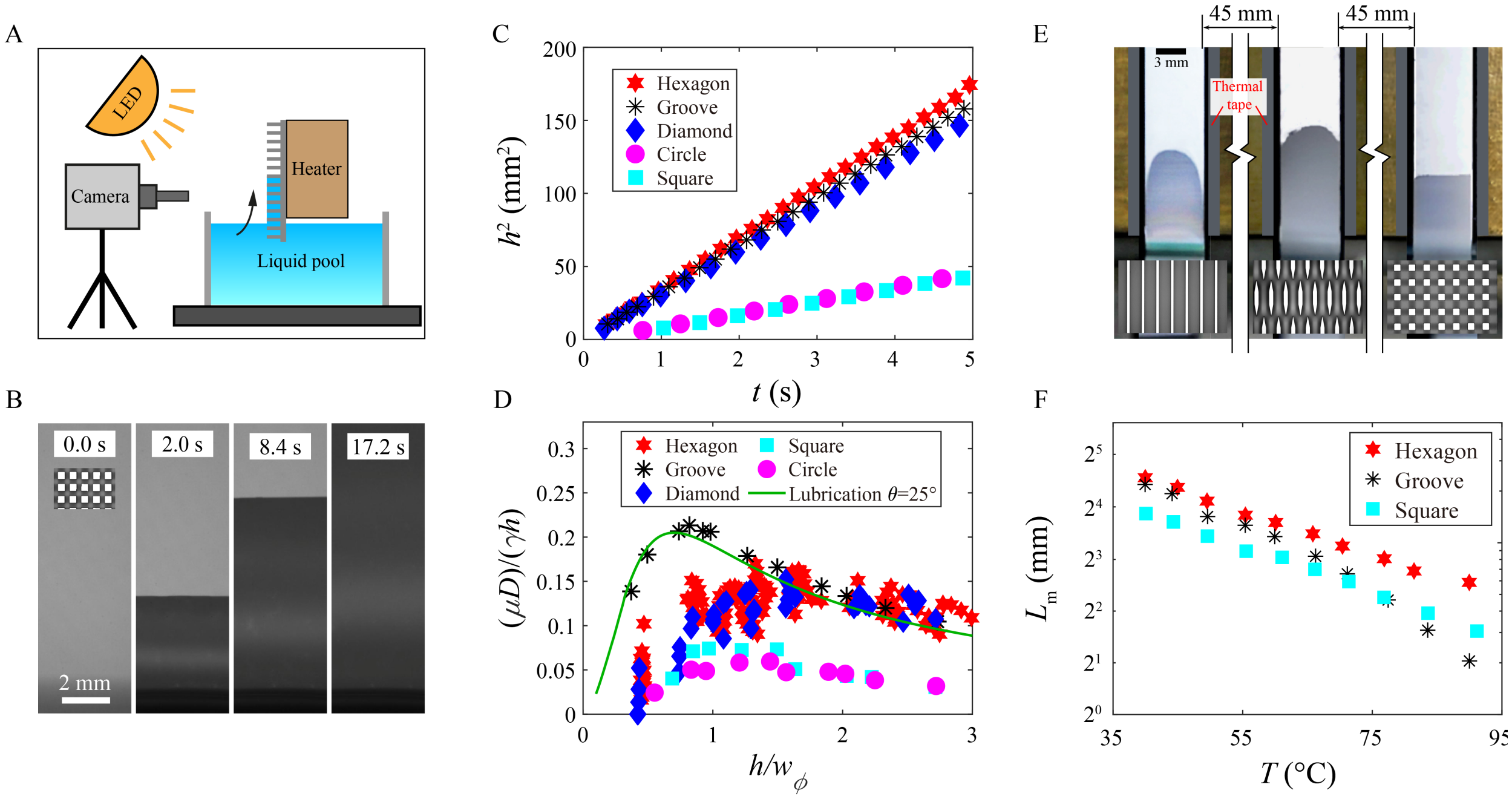

Fig. 4. Wicking coefficients of different textured surfaces and their wicking distances upon heating. A. Sketch of the capillary rise experiment. To extract the wicking coefficients, the heater is turned off and butanol is used as the liquid. For the heating experiments, ethanol is used as the liquid. B. Snapshots of a typical wicking process on a square pillar textured surface and its enlarged top view (inset). The pillar width and spacing are 9.7 and 6.3 µm, respectively. C. Square of the wicking distances $L$ as a function of time $t$ for five typical textured surfaces: staggered hexagon, staggered diamond, regular circle, regular square and square grooves. D. Dimensionless wicking coefficients $\mu D/(\gamma h)$ of all the surfaces plotted as a function of (equivalent) aspect ratio $h/w_\phi$, where $w_\phi$ is the width of the corresponding groove structure which maintains the same porosity $\phi$. For grooved surface, $w_\phi = w$. All the results are obtained with an equilibrium contact angle $\theta_e \approx 25°$ and the solid line is the theoretical prediction of a lubrication-based-model[38] at $\theta_e = 25°$. E. The wicking states of three textured surfaces put on a heating plate at $T = 66$°C. The three samples are (from left to right): grooves, staggered hexagon, regular square, and they are kept about 45 mm apart. Ethanol is used as the liquid and its equilibrium contact angle with the substrate is $\theta_e = 27°$. F. Average wicking distance $L_e$ of the three samples at different heating temperatures.

Apart from the critical contact angle for wicking, the wicking speed on textured surfaces is also important for many applications. For example, it has been found that the wicking speed has a strong influence on the capillary limit of heat pipes[6] and critical heat flux (CHF) of boiling heat transfer[8,36]. Normally, the wicking distance $L$ increases with time $t$ diffusively: $L^2 = Dt$, where $D$ is the wicking coefficient which characterizes the wicking speed.

To see how do the two new types of textured surface perform in terms of the wicking speed, their wicking coefficients in the fastest wicking direction are experimentally measured. Their geometrical parameters $b$, $c$, $e$ are varied to change the surface fraction $f$ and roughness $r$, while $a$ and $\alpha$ are kept constant. For comparison, the wicking coefficients of three more types of textured surfaces are also measured: (i) regularly arranged square pillars of length $a$ and height $h$; (ii) regularly arranged circular pillars of diameter $a$ and height $h$; and (iii) uniformly spaced parallel square grooves of width $w$ and depth $h$. The spacing between pillars of surfaces (i) and (ii) is designated as $b$, and $b$ or $w$ are varied to change the surface fraction $f$ and roughness $r$. All five types of textured surfaces are made of silicon and their height/depth $h$ are kept the same ($\approx 30$ μm) to facilitate comparison. See SM for details on the parameters of the surfaces. Butanol is used as the liquid and its equilibrium contact angle with the surfaces is measured to be around 25°. See SM for more details.

The wicking distances $L$ are measured by the capillary rise method, see Fig. 4A for a sketch of the experimental setup and Fig. 4B the snapshots of a typical experiment on square pillars. The wicking coefficients $D$ are extracted from the $L$ vs. $t$ curves (see Fig. 4C for five typical curves) and the nondimensionalized wicking coefficients $\mu D/(\gamma h)$ of all samples are plotted in Fig. 4D. Here, $\mu$ and $\gamma$ are the viscosity and surface tension of the liquid. Again, to facilitate comparison between pillar-textured surfaces (disconnected texture) with grooves (continuous texture), an equivalent width $w_\phi$ is used in the aspect ratio $h/w_\phi$, where $w_\phi$ is the width of the corresponding groove structure which maintains the same porosity $\phi$. Thus, the volume flux $\dot{V}$ of the wicking liquid becomes proportional to the wicking coefficient $D$ (see SM for details).

Solid line in Fig. 4D is the wicking coefficients of square grooves predicted by a lubrication-theory-based model[38], which fits well with the experimental results. Most of the surfaces with disconnected texture have smaller wicking coefficients than that of grooves. However, when $h/w_\phi > 1.5$, properly designed new surfaces could have larger wicking coefficients than that of grooves. This is surprising in the sense that grooves are supposed to have larger wicking coefficients than that of disconnected textures because grooves at least provide continuous driving force without the contact line pinning associated with discrete obstacles[17,18]. Consequently, this finding suggests that the new type of textured surface could have better heat transfer rates and CHF in liquid-vapor phase change heat exchangers. As a demonstration, capillary rise experiments with uniform heating at the backside of the textured silicon surfaces were performed with ethanol being the liquid. The three samples, staggered hexagon, regular square and grooves, have the same porosity $\phi = 0.8$ and (equivalent) aspect ratio $h/w_\phi = 2$. The stable wicking states at different heating temperatures $T$ are recorded (see Fig. 4E for an example at $T = 66$°C) and the average wicking length $L_m$ is plotted in Fig. 4F. Larger $L_m$ means stronger liquid pumping ability, which can reduce the size of the dry spot during boiling heat transfer[8,36], thus increasing CHF. It can be seen that at any heating temperature above 33°C, the staggered hexagon surface has the largest wicking length. Plus, $L_m$ of grooves and regular square surface decreases much faster than that of staggered hexagons surface, for example, at $T = 90$°C, $L_m$ of the staggered hexagon surface becomes ~ 2 times of the regular square surface and ~ 3 times of the grooved surface. This means that the newly designed staggered hexagon surfaces could have much better liquid pumping ability at higher temperatures and could lead to much larger CHF than traditional textured surfaces.

## References

1. Bell, J. M. & Cameron, F. K. The Flow of Liquids through Capillary Spaces. *J. Phys. Chem.* **10**, 658–674 (1906).
2. Washburn, E. W. The dynamics of capillary flow. *Phys. Rev.* **17**, 273–283 (1921).
3. De Gennes, P.-G., Brochard-Wyart, F. & Quéré, D. *Capillarity and Wetting Phenomena: Drops, Bubbles, Pearls, Waves*. (Springer Science & Business Media, 2003).
4. Raux, P. S., Cockenpot, H., Ramaioli, M., Quéré, D. & Clanet, C. Wicking in a Powder. *Langmuir* **29**, 3636–3644 (2013).
5. Nam, Y., Sharratt, S., Byon, C., Sung Jin Kim & Ju, Y. S. Fabrication and Characterization of the Capillary Performance of Superhydrophilic Cu Micropost Arrays. *J. Microelectromech. Syst.* **19**, 581–588 (2010).
6. Groll, M. Heat Pipe Science and Technology: A Historical Review. *Heat Pipe Sci. Technol. Int. J.* **5**, 1–58 (2014).
7. Cho, H. J., Preston, D. J., Zhu, Y. & Wang, E. N. Nanoengineered materials for liquid–vapour phase-change heat transfer. *Nat. Rev. Mater.* **2**, 16092 (2016).
8. Song, Y., Zhang, L., Díaz-Marín, C. D., Cruz, S. S. & Wang, E. N. Unified descriptor for enhanced critical heat flux during pool boiling of hemi-wicking surfaces. *Int. J. Heat Mass Transf.* **183**, 122189 (2022).
9. Martinez, A. W., Phillips, S. T., Butte, M. J. & Whitesides, G. M. Patterned Paper as a Platform for Inexpensive, Low-Volume, Portable Bioassays. *Angew Chem Int Ed* **46**, 1318–1320 (2007).
10. Gao, W. *et al.* Fully integrated wearable sensor arrays for multiplexed in situ perspiration analysis. *Nature* **529**, 509–514 (2016).
11. Xu, B. *et al.* Elastic Janus film for Wound Dressings: Unidirectional Biofluid Transport and Effectively Promoting Wound Healing. *Adv Funct Materials* **31**, 2105265 (2021).
12. Forny, L., Marabi, A. & Palzer, S. Wetting, disintegration and dissolution of agglomerated water soluble powders. *Powder Technol.* **206**, 72–78 (2011).
13. Duprat, C. Moisture in Textiles. *Annual Review of Fluid Mechanics* vol. 54 443–467 (2022).
14. Hettiarachchi, D., Michielsen, S., Wen, C. & Wang, L. Liquid moisture wicking in textile and inter-fibre pore filling behaviour in yarns: a review. *J. Text. Inst.* **116**, 2500–2519 (2025).
15. Mai, T. T. *et al.* Dynamics of Wicking in Silicon Nanopillars Fabricated with Interference Lithography and Metal-Assisted Chemical Etching. *Langmuir* **28**, 11465–11471 (2012).
16. Chu, K.-H., Xiao, R. & Wang, E. N. Uni-directional liquid spreading on asymmetric nanostructured surfaces. *Nat. Mater.* **9**, 413–417 (2010).
17. Bico, J., Tordeux, C. & Quéré, D. Rough wetting. *Europhys. Lett.* **55**, 214 (2001).
18. Courbin, L. *et al.* Imbibition by polygonal spreading on microdecorated surfaces. *Nat. Mater.* **6**, 661–664 (2007).

19. Cavalli, A., Blow, M. L. & Yeomans, J. M. Modelling unidirectional liquid spreading on slanted microposts. *Soft Matter* **9**, 6862 (2013).
20. Vrancken, R. J. *et al.* Anisotropic wetting and de-wetting of drops on substrates patterned with polygonal posts. *Soft Matter* **9**, 674–683 (2013).
21. Semprebon, C., Forsberg, P., Priest, C. & Brinkmann, M. Pinning and wicking in regular pillar arrays. *Soft Matter* **10**, 5739–5748 (2014).
22. Srivastava, N., Din, C., Judson, A., MacDonald, N. C. & Meinhart, C. D. A unified scaling model for flow through a lattice of microfabricated posts. *Lab Chip* **10**, 1148 (2010).
23. Extrand, C. W., Moon, S. I., Hall, P. & Schmidt, D. Superwetting of Structured Surfaces. *Langmuir* **23**, 8882–8890 (2007).
24. Jokinen, V., Leinikka, M. & Franssila, S. Microstructured Surfaces for Directional Wetting. *Adv. Mater.* **21**, 4835–4838 (2009).
25. Xiao, R., Enright, R. & Wang, E. N. Prediction and Optimization of Liquid Propagation in Micropillar Arrays. *Langmuir* **26**, 15070–15075 (2010).
26. Xiao, R. & Wang, E. N. Microscale Liquid Dynamics and the Effect on Macroscale Propagation in Pillar Arrays. *Langmuir* **27**, 10360–10364 (2011).
27. Kim, S. J., Kim, J., Moon, M.-W., Lee, K.-R. & Kim, H.-Y. Experimental study of drop spreading on textured superhydrophilic surfaces. *Phys. Fluids* **25**, 092110 (2013).
28. Kim, B. S., Lee, H., Shin, S., Choi, G. & Cho, H. H. Interfacial wicking dynamics and its impact on critical heat flux of boiling heat transfer. *Appl. Phys. Lett.* **105**, 191601 (2014).
29. Hisler, V. *et al.* Model Experimental Study of Scale Invariant Wetting Behaviors in Cassie–Baxter and Wenzel Regimes. *Langmuir* **30**, 9378–9383 (2014).
30. Liu, B. Y., Seemann, R., Chen, L. J. & Brinkmann, M. Directional liquid wicking in regular arrays of triangular posts. *Langmuir* **35**, 16476–16486 (2019).
31. Wang, C., Rahman, M. M. & Bucci, M. Decrypting the mechanisms of wicking and evaporation heat transfer on micro-pillars during the pool boiling of water using high-resolution infrared thermometry. *Phys. Fluids* **35**, 037112 (2023).
32. Zhang, S. *et al.* Prediction of hemiwicking dynamics in micropillar arrays. *Phys. Fluids* **35**, 082115 (2023).
33. Nam, H. T., Cho, H. H., Lee, S. & Lee, D. Temperature-dependent wicking dynamics and its effects on critical heat flux on micropillar structures in pool boiling heat transfer. *Int. Commun. Heat Mass Transf.* **146**, 106887 (2023).
34. Somlyai-Sipos, L. & Baumli, P. Wettability of Metals by Water. *Metals* **12**, 1274 (2022).
35. Hierro-Oliva, M., Gallardo-Moreno, A. M., Rodríguez-Cano, A., Bruque, J. M. & González-Martín, M. L. Experimental approach towards the water contact angle value on the biomaterial alloy Ti6Al4V. *Ann. Univ. Mariae Curie-Skłodowska, Sect. AA, Chem.* **70**, 1–13 (2015).
36. Dhillon, N. S., Buongiorno, J. & Varanasi, K. K. Critical heat flux maxima during boiling crisis on textured surfaces. *Nat. Commun.* **6**, 8247 (2015).
37. Mandel, R. K., Bae, D. G. & Ohadi, M. M. Embedded Two-Phase Cooling of High Flux Electronics Via Press-Fit and Bonded FEEDS Coolers. *J. Electron. Packag.* **140**, 031003 (2018).
38. Kolliopoulos, P. *et al.* Capillary-flow dynamics in open rectangular microchannels. *J. Fluid Mech.* **911**, A32 (2021).

**Acknowledgments:**

**Funding:** Y. L. acknowledges the financial support from the National Natural Science Foundation of China under grant No. 12272376.

**Author contributions:**

Conceptualization: YL

Data curation: ZL, CM, LCH, YL

Methodology: YL, CM

Investigation: ZL, CM, LCH, YL

Visualization: ZL, CM, LCH, YL

Funding acquisition: YL

Supervision: YL

Writing – original draft: YL, LCH

**Competing interests:** Y.L., Z.L. and C.M. are co-authors on a filed US patent no. 19/710,473, which describes the methods used herein.

**Data and materials availability:** All data are available in the main text or the supplementary materials.

## Supplementary Materials

Materials and Methods